\documentclass{IEEE_lsens}
\usepackage{textcomp}
\usepackage{xcolor}

\usepackage[noadjust]{cite}

\ifCLASSINFOpdf
\else
\fi
\usepackage[T1]{fontenc} 
\usepackage{amsmath}
\usepackage[cmintegrals]{newtxmath}
\usepackage{bm}
\usepackage{amsmath,amssymb}
\usepackage{amsfonts}
\usepackage{amsbsy}
\usepackage{graphicx}

\usepackage{subfigure}
\usepackage{epstopdf}
\usepackage{lipsum}
\usepackage{multirow}

\usepackage{array}
\usepackage{url}
\ifCLASSINFOpdf
\else
\fi
\providecommand{\hypersetup}[1]{\relax}

\begin{document}

\markboth{Vol.~PP, No.~99, May~2019}{0000000}

\IEEELSENSarticlesubject{Sensor Networks}

%
\title{Performance of Wireless Powered Cognitive Radio Sensor Networks with Nonlinear Energy Harvester}

%
\author{\IEEEauthorblockN{Devendra~S.~Gurjar\IEEEauthorrefmark{1}\IEEEauthorieeemembermark{1}, Ha~H.~Nguyen\IEEEauthorrefmark{2}\IEEEauthorieeemembermark{2},
and~Prabina~Pattanayak\IEEEauthorrefmark{1}\IEEEauthorieeemembermark{1}}
\IEEEauthorblockA{\IEEEauthorrefmark{1}Department of Electronics and Communication Engineering, National Institute of Technology, Silchar, 788010, India\\
\IEEEauthorrefmark{2}Department of Electrical and Computer Engineering, University of Saskatchewan,
Saskatoon, SK S7N 5A9, Canada\\
\IEEEauthorieeemembermark{1}Member, IEEE\\
\IEEEauthorieeemembermark{2}Senior Member, IEEE}
\thanks{Corresponding author: Devendra S. Gurjar (e-mail: devendra.gurjar@ieee.org).\protect\\}
\thanks{Digital Object Identifier 10.1109/LSENS.2019.0000000}}
%
%
%


\IEEEtitleabstractindextext{%
\begin{abstract}
This letter analyzes the performance of simultaneous wireless information-and-power transfer (SWIPT) in a cognitive radio sensor network (CRSN) under Nakagami-$m$ fading. A pair of sensor nodes (SNs) is considered in which one SN facilitates relay cooperation for communications between two primary users (PUs). In return, SNs make use of primary users' signals for energy harvesting (EH) and realize their own communications. In such a network,  bidirectional communications between the two PUs and unidirectional information exchange between the SNs can be performed in three phases. A power splitting (PS) based approach is adopted for enabling SWIPT. The relaying SN applies an amplify-and-forward (AF) protocol to broadcast primary and secondary signals, whereas the PUs perform selection combining to access the active direct link. Accurate expressions of the outage probability (OP) and throughput are derived for the primary system by considering a nonlinear energy harvester at the relaying SN under Nakagami-$m$ fading.
\end{abstract}

}


\maketitle

\section{Introduction}
Recent studies have demonstrated the applicability of (i) spectrum sharing techniques in improving spectrum utilization \cite{Khan2017}, and (ii) radio frequency (RF) based energy harvesting in prolonging the lifetime of wireless sensor networks (WSNs) \cite{Song2018,Varshney}. Due to increasing number of wireless devices, the industrial, scientific and medical (ISM) band is becoming congested day by day. As the conventional WSNs rely on ISM band, it is highly difficult and challenging to guarantee the required quality of service \cite{Du2018}. The problem of spectrum congestion can be solved by integrating the concept of cognitive radios network (CRN) with WSNs \cite{Ahmad}. As a result, cognitive radio sensor networks (CRSNs)  \cite{Bicen} can be very useful to realize a reliable as well as low-cost remote monitoring systems for many applications. Naturally, employing simultaneous wireless information and power transfer (SWIPT) and spectrum sharing in WSNs can tackle the two fundamental problems of spectrum scarcity and network lifetime. Specifically, three receiver designs, namely power splitting (PS), time switching, and antenna switching have been considered to exploit SWIPT in various wireless and mobile networks \cite{Nasir,Zhou}. Among these designs, PS-SWIPT is adopted in this paper, where the received power is split into two parts, one for information processing (IP) and another for energy harvesting (EH). On the other hand, to facilitate spectrum sharing, three approaches are commonly studied in the literature: interweave, underlay, and overlay. Among these, overlay-based spectrum sharing is of particular interest in this paper where both primary users (PUs) and sensor nodes (SNs) can transmit their signals in the same licensed band with the condition that the SNs provide relay cooperation to the PUs.

Several works have considered SWIPT in cooperative CRNs \cite{Im2015}--\cite{Nguyen2018}. Specifically, the authors in \cite{Im2015} and \cite{Yang2016} have considered an underlay CRN with RF energy harvesting and analyzed outage performance of the system. As an extension to the system in \cite{Im2015,Yang2016}, the authors in \cite{Kalamkar2017} have considered multiple primary nodes and derived expressions of the outage probability (OP) and ergodic capacity for the secondary system by considering interference from multiple PUs. Further, in \cite{Verma2017}, the authors have introduced one-way cooperative CRNs with SWIPT and analyzed the OP and throughput performance for the primary and secondary systems. Similar to \cite{Verma2017}, opportunistic relaying has been studied in \cite{Yan2017} by considering a dynamic SWIPT technique. Recently, the authors in \cite{Nguyen2018} have examined the outage performance of EH-enabled cooperative CRNs over Nakagami-$m$ fading environment. It should be pointed out, however, that all the above mentioned works have considered only the linear energy-harvesting model. Given that recent developments have shown that the linear EH model is not practical because the EH circuit is usually made up of diodes, capacitors, and inductors \cite{nonEH}. Therefore, the objective of this paper is to analyze performance of a bidirectional SWIPT-based CRSN equipped with a nonlinear energy harvester circuit.

Specifically, this paper adopts a piece-wise linear EH model to capture the saturation effect of a practical EH circuit as proposed in \cite{Dong},  \cite{Zhang2018}. A three-phase relaying protocol is considered to exploit the direct link, where one SN provides relay cooperation to two communicating PUs. In return, SNs utilize the primary transmission for EH and realizing their communications. An amplify-and-forward (AF) operation is considered at the relaying SN to broadcast primary and secondary signals, whereas selection combining (SC) is adopted at PUs to utilize the active direct link. For such a protocol, expressions of the user's OP and throughput of the primary system are obtained under Nakagami-$m$ fading. Numerical and simulation results enlighten the impacts of different system/channel parameters on the performance of SWIPT-enabled CRSNs.

{\textit{Notations}: $f_{X}(\cdot)$ and $F_{X}(\cdot)$ represent the probability density function (PDF) and the cumulative distribution function (CDF) of a random variable $X$, respectively, and $\textmd{Pr}[\cdot]$ denotes probability. $\Gamma[\cdot,\cdot]$, $\Upsilon[\cdot,\cdot]$, and $\Gamma[\cdot]$ represent, respectively, the upper incomplete, the lower incomplete, and the complete Gamma functions \cite[eq. (8.350)]{math}.}
\vspace*{-0.15cm}
\section{System and Protocol Description}\label{sysmod}
\begin{figure}[t]
	\centering
	\includegraphics[width=2.5in]{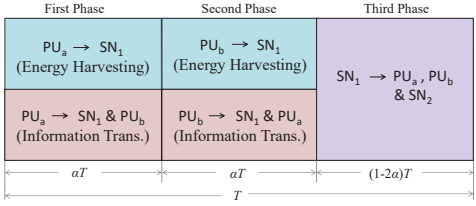}
	\caption{Signaling in the three-phase CRSN with PS-SWIPT.}
	\label{fig1}
\end{figure}

This paper considers a three-phase SWIPT-based CRSN where two primary users ${\sf PU}_{a}$ and ${\sf PU}_{b}$ communicate to each other with the help of a SWIPT-enabled node ${\sf SN}_{1}$. All PUs and SNs operate in a half-duplex mode and each is equipped with a single antenna. Fig. \ref{fig1} depicts the involved operations in one communication block for the considered system. In particular, three-time phases are required to accomplish the end-to-end information exchange between two PUs. In the first phase, ${\sf PU}_{a}$ transmits a signal to ${\sf SN}_{1}$ and ${\sf PU}_{b}$. Likewise, in the second phase, ${\sf PU}_{b}$ transmits a signal to ${\sf SN}_{1}$ and ${\sf PU}_{a}$. A PS strategy is employed at ${\sf SN}_{1}$ to harvest energy from RF signals received in the first two consecutive phases from PUs. In the third phase, ${\sf SN}_{1}$ applies an AF operation to broadcast the combined primary signal after adding its own signal intended for ${\sf SN}_{2}$.

The channel coefficients of all the communication links are affected by quasi-static fading. As such, for one communication block (three-time phases), they remain unchanged. Further, channel coefficients corresponding to ${\sf PU}_{a}$--${\sf SN}_{1}$ and ${\sf SN}_{1}$--${\sf PU}_{b}$ links are represented as $h_{a,1}$ and $h_{1,b}$, respectively. Similarly, channel coefficients corresponding to ${\sf PU}_{a}$--${\sf PU}_{b}$ and ${\sf SN}_{1}$--${\sf SN}_{2}$ links are denoted as $h_{a,b}$ and $h_{1,2}$. Moreover, all the channel coefficients are considered to be reciprocal and follow Nakagami-$m$ distributions.
\vspace*{-0.15cm}
\subsection{Energy Harvesting}
In the first phase,  let ${\sf PU}_{a}$ transmits a symbol $x_{a}$. Then, the signals received at ${\sf SN}_{1}$ and ${\sf PU}_{b}$ can be given, respectively, as $y^{\rm (I)}_{a,1}=\sqrt{P_{a}}h_{a,1}x_{a}+n^{\rm (I)}_{1}$ and $y^{\rm (I)}_{a,b}=\sqrt{P_{a}}h_{a,b}x_{a}+n^{\rm (I)}_{b}$, where $x_a$ is normalized to unit power whereas $P_{a}$ denotes the actual transmit power at ${\sf PU}_{a}$. Further, $n^{\rm (I)}_{1}\sim\mathcal{CN}(0,\sigma^{2}_{1})$ and $n^{\rm (I)}_{b}\sim\mathcal{CN}(0,\sigma^{2}_{b})$ represent additive white Gaussian noise (AWGN) variables at ${\sf SN}_{1}$ and ${\sf PU}_{b}$, respectively. Likewise, in the second phase, the received signals at ${\sf SN}_{1}$ and ${\sf PU}_{a}$ are given, respectively, as $y^{\rm (II)}_{b,1}=\sqrt{P_{b}}h_{b,1}x_{b}+n^{\rm (II)}_{1}$ and $y^{\rm (II)}_{b,a}=\sqrt{P_{b}}h_{b,a}x_{b}+n^{\rm (II)}_{a}$, where $P_{b}$ is the transmit power, $n^{\rm (II)}_{1}\sim\mathcal{CN}(0,\sigma^{2}_{1})$ and $n^{\rm (II)}_{a}\sim\mathcal{CN}(0,\sigma^{2}_{a})$ are AWGN variables at ${\sf SN}_{1}$ and ${\sf PU}_{a}$, respectively.
During the first and second phases, ${\sf SN}_{1}$ uses a fraction of signals $\sqrt{\beta} y^{\rm (I)}_{a,1}$ and $\sqrt{\beta} y^{\rm (II)}_{b,1}$ to harvest energy from the received signals, where $0<\beta<1$. Hereafter, ${\sf SN}_{1}$ stores this energy using an appropriate circuit and makes use of it in the third phase for applying an AF operation on the combined signals. The remaining fractions of the signals, namely $\sqrt{1-\beta} y^{\rm (I)}_{a,1}$ and $\sqrt{1-\beta} y^{\rm (II)}_{b,1}$, are allocated for information processing and broadcasting.

Further, the harvested energy at ${\sf SN}_{1}$ in the first phase is expressed as $\mathcal{E}^{\rm (I)}_{h}=\eta\beta P_{a} |h_{a,1}|^{2}\alpha T$, where $0<\eta<1$ denotes the energy conversion efficiency of the circuit in the linear region. Similarly, the harvested energy at ${\sf SN}_{1}$ in the second phase is given as $\mathcal{E}^{\rm (II)}_{h}=\eta\beta P_{b} |h_{b,1}|^{2}\alpha T$. The total harvested energy at ${\sf SN}_{1}$ is thus $\mathcal{E}_{h}=\eta\beta (P_{a} |h_{a,1}|^{2}+P_{b} |h_{b,1}|^{2})\alpha T$.

Considering the nonlinear EH model as in \cite{Dong}, the transmit power at ${\sf SN}_{1}$ can be formulated as
\begin{align}\label{sdgee}
\!\!\!\!P_{1}\!=\!\left\{\!\!\!\begin{array}{l}
\frac{\!\!\!\alpha\eta\beta (P_{a} |h_{a,1}|^{2}\!+\!P_{b} |h_{b,1}|^{2})}{1-2\alpha},\,\,\,\,P_{a} |h_{a,1}|^{2}\!+\!P_{b} |h_{b,1}|^{2}\leq P_{\textmd{th}} \\
\frac{\!\!\!\alpha \eta \beta P_{\textmd{th}}}{1-2\alpha},\qquad \quad\quad\quad \,\,\,\,\,\,\,\,\,P_{a} |h_{a,1}|^{2}\!+\!P_{b} |h_{b,1}|^{2}>P_{\textmd{th}}
\end{array}\right.
\end{align}
where $P_{\textmd{th}}$ is the saturation threshold of the EH circuit\footnote{It is assumed that the energy consumption in processing of information signals is negligible and the total harvested energy is being utilized for signal broadcasting in the third phase \cite{Nasir}}.

\vspace*{-0.15cm}
\subsection{Signal-to-Noise Ratio (SNR)}
After splitting the signals received in the first two phases, ${\sf SN}_{1}$ combines two signals $\sqrt{1-\beta}y^{\rm (I)}_{a,1}$ and $\sqrt{1-\beta}y^{\rm (II)}_{b,1}$ by applying an AF operation with gain $\mathcal{G}\approx\sqrt{\mu P_{1}/((1-\beta)(P_{a}|h_{a,1}|^{2}+P_{b}|h_{b,1}|^{2}))}$ \cite{Lei}. The transmit signal in the third phase, i.e., the broadcasting phase, can be expressed as
{\small{\begin{align}\label{dwuwuh}
		\!\!x^{\textmd{(BC)}}_{1}&\!=\!\mathcal{G}\big(\!\sqrt{(1\!-\!\beta)P_{a}}h_{a,1}x_{a}\!+\!\sqrt{1\!-\!\beta}n^{\rm (I)}_{1}\!+\!n^{\rm (I)}_{\rm cr}
		\!\!+\!\sqrt{(1\!-\!\beta)P_{b}}h_{b,1}x_{b}\nonumber\\
		&+\!\sqrt{1\!-\!\beta}n^{\rm (II)}_{1}\!+\!n^{\rm (II)}_{\rm cr}\big)
		\!+\!\sqrt{\!(1-\mu)P_{1}}x_{1}
		\end{align}}}
where $n^{\rm (I)}_{\rm cr}\sim\mathcal{CN}(0,\sigma^{2}_{\rm cr})$ and $n^{\rm (II)}_{\rm cr}\sim\mathcal{CN}(0,\sigma^{2}_{\rm cr})$ are noise components generated in the RF-to-baseband conversion processes in the first and second phases, respectively, and $\mu\in(0,1]$ is power allocation factor at the relaying SN. The signal received at PUs in the third phase can be expressed as
\begin{align}\label{sdeew}
y^{\rm (III)}_{1,j}=h_{1,j}x^{\textmd{(BC)}}_{1}+n^{\rm (III)}_{j}
\end{align}
where $j\in\{a,b\}$ and $n^{\rm (III)}_{j}\sim\mathcal{CN}(0,\sigma^{2}_{j})$.

Since both ${\sf PU}_{a}$ and ${\sf PU}_{b}$ know their own transmitted signals, they can cancel the self-interference. After performing self-interference cancellation, the instantaneous
SNR concerning the transmission from ${\sf SN}_{1}$ to ${\sf PU}_{j}$ in the third phase can be expressed as
\begin{align}\label{sdggee}
\gamma_{1,j}=\left\{\!\! \begin{array}{l}
\gamma^{\textmd{(lin)}}_{1,j},\quad P_{a} |h_{a,1}|^{2}+P_{b} |h_{b,1}|^{2}\leq P_{\textmd{th}} \\
\gamma^{\textmd{(sat)}}_{1,j},\quad P_{a} |h_{a,1}|^{2}+P_{b} |h_{b,1}|^{2}>P_{\textmd{th}}
\end{array}\right.
\end{align}
where $\gamma^{\textmd{(lin)}}_{1,j}$ and $\gamma^{\textmd{(sat)}}_{1,j}$ are given for $j,\hat{j}\in\{a,b\}$, $\hat{j}\neq j$, as
\begin{align}\label{gkhj}
\gamma^{\textmd{(lin)}}_{1,j}=\frac{\varepsilon_{1}|h_{\hat{j},1}|^{2}}{\varepsilon_{2}+\varepsilon_{3}|h_{1,j}|^{2}+\varepsilon_{4}|h_{\hat{j},1}|^{2}}
\end{align}
and
{\small{\begin{align}\label{gkshj}
		 \!\gamma^{\textmd{(sat)}}_{1,j}\!=\!\frac{\varphi_{1}|h_{\hat{j},1}|^{2}|h_{1,j}|^{2}}{\varphi_{2}|h_{1,j}|^{2}\!+\!\varphi_{3}|h_{1,j}|^{4}\!+\!\varphi_{4}|h_{1,j}|^{2}|h_{\hat{j},1}|^{2}\!+\!\varphi_{5}|h_{\hat{j},1}|^{2}}
		\end{align}}}
where $\varepsilon_{1}=\mu P_{\hat{j}}$, $\varepsilon_{2}=2\mu \sigma^{2}_{1}+\frac{2\mu\sigma^{2}_{\rm cr}}{1-\beta}$, $\varepsilon_{3}=P_{j}(1-\mu)$, $\varepsilon_{4}=P_{\hat{j}}(1-\mu)$, $\varphi_{1}=\mu\delta P_{\textmd{th}}P_{\hat{j}}$, $\varphi_{2}=2\mu\delta P_{\textmd{th}}\sigma^{2}_{1}+\frac{2\mu\delta P_{\textmd{th}}\sigma^{2}_{\rm cr}}{1-\beta}+P_{j}\sigma^{2}_{j}$, $\varphi_{3}=P_{j}(1-\mu)P_{\textmd{th}}\delta$, $\varphi_{4}=P_{\hat{j}}(1-\mu)P_{\textmd{th}}\delta$, $\varphi_{5}=P_{\hat{j}}\sigma^{2}_{j}$, and $\delta=\alpha\eta\beta/(1-2\alpha)$.
In obtaining the expression for $\gamma^{\textmd{(lin)}}_{1,j}$, the effect of noise terms in the third phase at PUs is considered negligible compared to noise due to RF-to-baseband conversion \cite{HLee} and interference due to the secondary transmission.


\section{Performance Analysis}
\setlength{\abovedisplayskip}{0pt}
\setlength{\belowdisplayskip}{0pt}
\subsection{Outage Probability Analysis}\label{opa}
For a delay-limited wireless system, OP is a critical performance metric that defines the probability of link failure. For the system considered in this paper, an outage event happens at the PUs if the instantaneous data rate (achieved by utilizing both direct and relayed transmissions) at the corresponding PU falls below a predefined target rate. Therefore, the user OP at ${\sf PU}_{j}$ is mathematically defined for $j,\hat{j}\in\{a,b\}$, $\hat{j}\neq j$, as
\begin{align}\label{oop}
\mathcal{P}_{\textmd{out},j}&\!=\!\textmd{Pr}[\mathcal{R}^{\textmd{SC}}_{\jmath,i}<r_{j}]
\!=\!\textmd{Pr}\left[\max\left(\gamma_{1,j},\gamma_{\hat{j},j}\right)<\bar{\gamma}_{j}\right]
\end{align}
where $r_{j}$ is the target rate and $\bar{\gamma}_{j}=2^{r_{j}/(1-2\alpha)}-1$ with $\alpha\in(0,0.5)$.
Since \eqref{oop} can be computed as
\begin{align}\label{dww}
\mathcal{P}_{\textmd{out},j}&=F_{\gamma_{1,j}}(\bar{\gamma}_{j}) F_{\gamma_{\hat{j},j}}(\bar{\gamma}_{j}).
\end{align}
one needs to obtain $F_{\gamma_{1,j}}(\bar{\gamma}_{j})$ for two cases of $\gamma_{1,j}$ given in \eqref{sdggee}. For the case when $P_{a} |h_{a,1}|^{2}+P_{b} |h_{b,1}|^{2}\leq P_{\textmd{th}}$, the expression of $F_{\gamma_{1,j}}(\bar{\gamma}_{j})$ is given in Lemma 1.
\newtheorem{lemma}{Lemma}
\begin{lemma}\label{lem1}
	The expression of $F_{\gamma^{\textmd{(lin)}}_{1,j}}(\bar{\gamma}_{j})$ is
	{\small{\begin{align}\label{saljssd}
			\!\!F_{\gamma^{\textmd{(lin)}}_{1,j}}(\bar{\gamma}_{j}) &\!=\!\frac{\Upsilon\left[m_{j},\frac{m_{j}P_{\textmd{th}}}{\Omega_{j}P_{j}}\right]}{\Gamma[m_{j}]}\!-\!\frac{\left(\frac{m_{j}}{\Omega_{j}}\right)^{m_{j}}{\rm e}^{-\frac{m_{\hat{j}}P_{\textmd{th}}}
					 {\Omega_{\hat{j}}P_{\hat{j}}}}}{\Gamma[m_{j}]}\!\!\!\sum^{m_{\hat{j}}-1}_{k=0}\frac{\left(\frac{m_{\hat{j}}}{\Omega_{\hat{j}}P_{\hat{j}}}\right)^{k}}{k!}\nonumber\\
			&\times \sum^{k}_{q=0}\binom{k}{q}P^{q}_{\textmd{th}}(-P_{j})^{k-q}\sum^{\nu}_{t=0}\frac{(-1)^{t}t!\binom{\nu}{t}}{\Xi^{t+1}}\nonumber\\
			&\times\Bigg\{{\rm e}^{\frac{\Xi P_{\textmd{th}}}{P_{j}}}\left(\frac{P_{\textmd{th}}}{P_{j}}\right)^{\nu-t}\!\!\!\!-{\rm e}^{\Xi\Delta_{1}}\Delta_{1}^{\nu-t}\Bigg\}-\frac{\left(\frac{m_{j}}{\Omega_{j}}\right)^{m_{j}}}{\Gamma[m_{j}]}\nonumber\\
			&\times {\rm e}^{-\frac{m_{\hat{j}}\bar{\gamma}_{j}\varepsilon_{2}}{\Omega_{\hat{j}}\Theta}}\sum^{m_{\hat{j}}-1}_{p=0}\frac{\left(\frac{m_{\hat{j}}}{\Omega_{\hat{j}}\Theta}\right)^{p}}{p!}\sum^{p}_{n=0}
			\binom{p}{n}\gamma^{p}_{\textmd{th}}\varepsilon^{n}_{2}\varepsilon^{p-n}_{3}\nonumber\\
			&\times\Upsilon\left[m_{j}+p-n,\Psi\Delta_{1}\right]\Psi^{-m_{j}-p+n}
			\end{align}}}
	where $\Theta=\varepsilon_{1}-\varepsilon_{4}\bar{\gamma}_{j}$, $\nu=m_{j}+k-q-1$, $\Xi=({m_{\hat{j}}P_{j}}/{\Omega_{\hat{j}}P_{\hat{j}}}-{m_{j}}/{\Omega_{j}})$, $\Xi\neq 0$, $\Psi=({m_{j}}/{\Omega_{j}}+{m_{\hat{j}}\bar{\gamma}_{j}\varepsilon_{3}}/{\Omega_{\hat{j}}\Theta})$, $\Delta_{1}=({P_{\textmd{th}}\Theta-\bar{\gamma}_{j}\varepsilon_{2}P_{\hat{j}}})/({\bar{\gamma}_{j}\varepsilon_{3}P_{\hat{j}}+P_{j}\Theta})$.
\end{lemma}
\begin{IEEEproof}
	Please see Appendix A.
\end{IEEEproof}

On the other hand, for the case when $P_{a} |h_{a,1}|^{2}+P_{b} |h_{b,1}|^{2}> P_{\textmd{th}}$, the expression of $F_{\gamma_{1,j}}(\bar{\gamma}_{j})$ is given in Lemma 2.
\begin{lemma}\label{lem2}
	The expression of $F_{\gamma^{\textmd{(sat)}}_{1,j}}(\bar{\gamma}_{j})$ is
	{\small{\begin{align}\label{saljd}
			\!\!\!F_{\gamma^{\textmd{(lin)}}_{1,j}}(\bar{\gamma}_{j}) &\!\!=\!\!\sum^{m_{j}-1}_{k=0}\frac{1}{k!}\left(\frac{m_{j}}{\Omega_{j}\Lambda_{2}}\right)^{k}\sum^{k}_{q=0}\binom{k}{q}(-\Lambda_{1})^{k-q}
			{\rm e}^{\frac{m_{j}\Lambda_{1}}{\Omega_{j}\Lambda_{2}}}\nonumber\\
			&\times\frac{\left(\frac{m_{\hat{j}}}{\Omega_{\hat{j}}}\right)^{m_{\hat{j}}}}{\Gamma[m_{\hat{j}}]}\Gamma\Bigg[m_{\hat{j}}+q, \left(\frac{m_{\hat{j}}}{\Omega_{\hat{j}}}+\frac{m_{j}}{\Omega_{j}\Lambda_{2}}\right)\Delta_{2}\Bigg]\nonumber\\
			&\times\left(\frac{m_{\hat{j}}}{\Omega_{\hat{j}}}\!+\!\frac{m_{j}}{\Omega_{j}\Lambda_{2}}\right)^{\!-m_{\hat{j}}\!-q}
			\!+\!\!\!\!\sum^{m_{j}-1}_{p=0}\!\frac{1}{p!}\!\left(\frac{m_{j}}{\Omega_{j}P_{j}}\right)^{p}
			\!\!\!\sum^{p}_{n=0}\!\!\binom{p}{n}\nonumber\\
			 &\times\frac{P^{n}_{\textmd{th}}\!(-P_{\hat{j}})^{p\!-\!n}}{\Gamma[m_{\hat{j}}]}\left(\frac{m_{\hat{j}}}{\Omega_{\hat{j}}}\right)^{m_{\hat{j}}}\!\!\!{\rm e}^{-\frac{m_{j}P_{\textmd{th}}}{\Omega_{j}P_{j}}}
			\!\!\sum^{\infty}_{s=0}\left(\frac{m_{j}P_{\hat{j}}}{\Omega_{j}P_{j}}\right)^{s}\nonumber\\
			&\times\frac{\Upsilon\left[m_{\hat{j}}+p+s-n,\frac{m_{\hat{j}}}{\Omega_{\hat{j}}}\Delta_{2}\right]}{s!}
			\left(\frac{m_{\hat{j}}}{\Omega_{\hat{j}}}\!\right)^{\!-\!m_{\hat{j}}\!-\!p\!-\!s\!+\!n}
			\end{align}}}
	where $\Lambda_{1}=\varphi_{2}\bar{\gamma}_{j}/(\varphi_{1}-\varphi_{4}\bar{\gamma}_{j})$, $\Lambda_{2}=\varphi_{3}\bar{\gamma}_{j}/(\varphi_{1}-\varphi_{4}\bar{\gamma}_{j})$, $\Delta_{2}=(\Lambda_{2}P_{\textmd{th}}+\Lambda_{1}P_{j})/(P_{j}+P_{\hat{j}}\Lambda_{2})$.
\end{lemma}
\begin{IEEEproof}
	Please see Appendix B.
\end{IEEEproof}
\vspace*{-0.10cm}
\subsection{System Throughput}
The throughput of the primary system is defined as the sum of average target rates at both PUs that can be attained in delay-limited transmission scenario over fading channels. With the OP expressions derived in Section \ref{opa}, the throughput is calculated as
\begin{align}\label{throughput}
\mathcal{S}_{\mathcal{P}}=(1-2\alpha)\big[(1-\mathcal{P}_{\textmd{out},a})r_{a}+(1-\mathcal{P}_{\textmd{out},b})r_{b}\big]
\end{align}
where $\mathcal{P}_{\textmd{out},a}$ and $\mathcal{P}_{\textmd{out},b}$ denote the respective OPs for the ${\sf PU}_{b}\rightarrow{\sf PU}_{a}$ and ${\sf PU}_{a}\rightarrow{\sf PU}_{b}$ links, computed as in \eqref{dww}.

\section{Numerical and Simulation Results}
\begin{figure}
	\centering
	\includegraphics[width=2.8in]{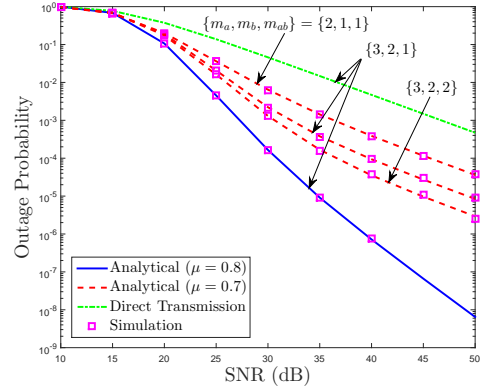}
	\caption{Outage probability versus SNR curves for the ${\sf PU}_{b}\rightarrow{\sf PU}_{a}$ link.}
	\label{fig2}
\end{figure}

Throughout this section, it is assumed that $P_{a}=P_{b}=P$, $\sigma^{2}_{j}=\sigma^{2}_{1}=\sigma^{2}_{\rm cr}=\sigma^{2}$, and the SNR is defined $\frac{P}{\sigma^{2}}$. Further, ${\sf PU}_{a}$ and ${\sf PU}_{b}$ are located at coordinates $(0,0)$ and $(8,0)$, respectively, whereas ${\sf SN}_{a}$ and ${\sf PU}_{b}$ are located at $(d,0)$ and $(d,4)$. Considering a linear relay network, the path-loss models are defined with $\Omega_{a}=d^{-v}$ and $\Omega_{b}=(8-d)^{-v}$, where $v=2.5$. The energy conversion efficiency and saturation threshold of the EH circuit are set as $0.7$ and $0$ dBm, respectively. The noise power at both PUs and SN is $-40$ dBm.

For the numerical results presented in Fig. \ref{fig2}, other system parameters are set as $d=4$, $\alpha=0.2$, $r_{a}=r_{b}=1/6$. This figure plots OP versus SNR curves by considering different values of fading severity parameters and power allocation factor. The OP performance of the considered SWIPT-CRSN is also compared with performance of direct transmission to show the advantage of relaying and energy harvesting. From Fig. \ref{fig2}, one can clearly see that as the value of fading severity parameters increases, the primary system enjoys improved OP performance. Further, one can also see that as the value of $\mu$ increases, the OP of the primary system improves significantly. This performance behavior is intuitively satisfying since increasing the value of $\mu$ reflects that more resource in terms of power is allocated for primary transmissions as compared to the secondary system.

\begin{figure}
	\centering
	\includegraphics[width=2.8in]{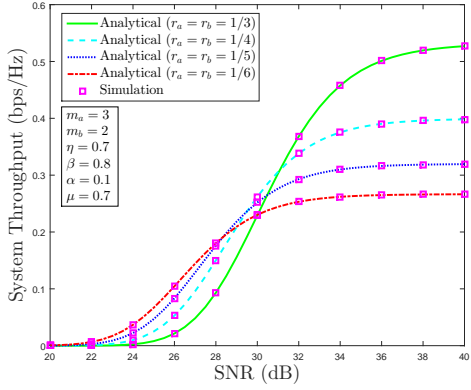}
	\caption{Throughput versus SNR curves for the primary system.}
	\label{fig3}
\end{figure}

Fig. \ref{fig3} demonstrates the primary system's throughput versus SNR for different target rates. Here, the parameters are set as $m_{a}=3, m_{b}=2$, $\beta=0.8$, $\alpha=0.1$ and $\eta=0.7$. From this figure, one can see that for low SNR values, the curves corresponding to a higher target rate exhibit a lower throughput.
The reason for this behavior is that as the target rate increases in the low SNR region, the corresponding target SNR also increases, which leads to degradation of the OP performance. When the OP of the primary system gets higher, the throughput performance degrades. In contrast, in the medium and high SNR regions, the degradation of OP performance is very small as compared to the throughput improvement when the target rates increase. Moreover, the system throughput curves become saturated at certain SNR values, which can be considered as the maximum achievable throughput values for the corresponding target rates.

\vspace*{-0.25cm}
\section{Conclusion}
This letter analyzed the performance of a popular three-phase cooperative SWIPT-enabled CRSN with a practical nonlinear energy harvester. First, outage probability expressions are derived for primary users under Nakagami-$m$ fading. Then, the obtained OP expressions are used to study the primary system's throughput. Numerical results validated the accuracy of the derived expressions and highlighted the effects of critical system and channel parameters on the system performance. It is evident from the numerical results that better channel conditions reflect better system performance in terms of both outage probability and system throughput. It is noteworthy that the considered system model can serve as a point of reference for several future works. For instance, it would be interesting to analyze the performance of the considered system in terms of bit error rate for different scenarios.


\vspace*{-0.25cm}
\appendices
\section{}\label{appA}
Let $X\triangleq|h_{1,j}|^{2}$, $Y\triangleq|h_{\hat{j},1}|^{2}$  for $j,\hat{j}\in\{a,b\}$ with $j\neq\hat{j}$. Under Nakagami-$m$ fading, $X$ and $Y$ follow the Gamma distribution with PDFs $f_{X}(x)=\left({m_{j}}/{\Omega_{j}}\right)^{m_{j}} ({1}/{\Gamma[m_{j}]})x^{m_{j}-1}{\rm e}^{-\frac{m_{j}x}{\Omega_{j}}},\,x\geq0$, $f_{Y}(y)=\left({m_{\hat{j}}}/{\Omega_{\hat{j}}}\right)^{m_{\hat{j}}} ({1}/{\Gamma[m_{\hat{j}}]})y^{m_{\hat{j}}-1}{\rm e}^{-\frac{m_{\hat{j}}y}{\Omega_{\hat{j}}}},\,y\geq0$.
On utilizing \eqref{sdggee}, the CDF $F_{\gamma^{\textmd{(lin)}}_{1,j}}(\bar{\gamma}_{j})$ can be expressed as
{\small{\begin{align}\label{djadja}
		 F_{\gamma^{\textmd{(lin)}}_{1,j}}(\bar{\gamma}_{j})&=\textmd{Pr}\left[\frac{\varepsilon_{1}Y}{\varepsilon_{2}+\varepsilon_{3}X+\varepsilon_{4}Y}<\bar{\gamma}_{j},
		P_{j}X+P_{\hat{j}}Y\leq P_{\textmd{th}}\right]\nonumber\\
		&=\textmd{Pr}\left[Y<\frac{\bar{\gamma}_{j}}{\Theta}(\varepsilon_{2}+\varepsilon_{3}X), Y\leq\frac{P_{\textmd{th}}-P_{j}X}{P_{\hat{j}}}\right].
		\end{align}}}
For $\Theta\geq 0$, \eqref{djadja} can be expressed in an integral form as
{\small{\begin{align}\label{ddja}
		 F_{\gamma^{\textmd{(lin)}}_{1,j}}(\bar{\gamma}_{j})&=\int^{\frac{P_{\textmd{th}}}{P_{j}}}_{\Delta_{1}}f_{X}(x)\int^{\frac{P_{\textmd{th}}-P_{j}x}{P_{\hat{j}}}}_{0}f_{Y}(y)dydx\nonumber\\
		&+\int^{\Delta_{1}}_{0}f_{X}(x)\int^{\frac{\bar{\gamma}_{j}}{\Theta}(\varepsilon_{2}+\varepsilon_{3}x)}f_{Y}(y)dydx
		\end{align}}}
where $\Theta$ and $\Delta_{1}$ are defined after \eqref{saljssd}. After applying some mathematical formulations and utilizing \cite[eqs. 3.381, 8.350]{math}, one can obtain
the expression given in Lemma \ref{lem1}. For the case when $\Theta<0$, $F_{\gamma^{\textmd{(lin)}}_{1,j}}(\bar{\gamma}_{j})=1$.
\vspace*{-0.25cm}
\section{}\label{appB}
The CDF $F_{\gamma^{\textmd{(sat)}}_{1,j}}(\bar{\gamma}_{j})$ can be expressed by utilizing \eqref{sdggee} as
{\small{\begin{align}\label{codh}
		\!\!\!\!F_{\gamma^{\textmd{(sat)}}_{1,j}}\!(\bar{\gamma}_{j})\!&=\!\textmd{Pr}\!\Bigg[Y\!<\!\!\frac{\varphi_{2}X\bar{\gamma}_{j}\!+\!\varphi_{3}
			\bar{\gamma}_{j}X^{2}}{\varphi_{1}X\!-\!\varphi_{4}\bar{\gamma}_{j}X\!-\!\varphi_{5}\bar{\gamma}_{j}},
		P_{j}X\!+\!P_{\hat{j}}Y\!>\!P_{\textmd{th}}\!\Bigg]
		\end{align}}}
Observe that in the medium to high SNR region $\varphi_{5}\bar{\gamma}_{j}$ is negligible as compared to other terms in the denominator. Therefore, \eqref{codh} can be expressed as
\begin{align}\label{csodh}
\!F_{\gamma^{\textmd{(sat)}}_{1,j}}\!(\bar{\gamma}_{j})&=\textmd{Pr}\!\left[X>\frac{Y-\Lambda_{1}}{\Lambda_{2}},X>\frac{P_{\textmd{th}}
	-P_{\hat{j}}Y}{P_{j}}\right]
\end{align}
where $\Lambda_{1}$ and $\Lambda_{2}$ are defined after \eqref{saljd}. For $\varphi_{1}-\varphi_{4}\bar{\gamma}_{j}\geq0$, $F_{\gamma^{\textmd{(sat)}}_{1,j}}\!(\bar{\gamma}_{j})$ can be written in an integration form as
{\small{\begin{align}\label{dfksk}
		F_{\gamma^{\textmd{(sat)}}_{1,j}}\!(\bar{\gamma}_{j})&=\int^{\infty}_{\Delta_{2}}f_{Y}(y)\int^{\infty}_{\frac{y-\Lambda_{1}}{\Lambda_{2}}}f_{X}
		(x)dxdy\nonumber\\
		&+\int^{\Delta_{2}}_{0}f_{Y}(y)\int^{\infty}_{\frac{P_{\textmd{th}}-P_{\hat{j}}y}{P_{j}}}f_{X}(x)dxdy
		\end{align}}}
Solving \eqref{dfksk} with the help of \cite[eqs. 3.381, 8.352]{math}, the final result is given in Lemma \ref{lem2}. For $\varphi_{1}-\varphi_{4}\bar{\gamma}_{j}<0$, $F_{\gamma^{\textmd{(sat)}}_{1,j}}\!(\bar{\gamma}_{j})=1$.

\end{document}